\documentclass[12pt]{article}
\usepackage{amsmath,amsfonts}

\DeclareMathOperator{\tr}{tr}

\begin{document}

\begin{center}
{\Large Scenario for the renormalization\\
in the 4D Yang-Mills theory}\\
\vspace{1cm}
{L.~D.~Faddeev}\\

\emph{\small St.~Petersburg dept. of Steklov}\\
\emph{\small Mathematical Institute}\\
\emph{\small and St.~Petersburg University}
\end{center}


    Yang-Mills quantum field theory has unique character, allowing a
    selfconsistent formulation in the four dimensional space-time.
    Two important properties --- asymptotic freedom and dimensional
    transmutation --- are characteristic features of this theory, which
    distinguish it from others. I think, that the typical textbook exposition
    of this theory, based on general paradigm of QFT, still does not
    underline these specifics. In my talk I propose a scheme for the
    description of the Yang-Mills theory which exactly does this.
    I do not claim finding anything new, my proposal has simply
    a methodological value.

    As a main tool for my presentation I have chosen the object, called
    the effective action, defined via the background field.
    Following Feynman ideas I consider the functional of a background
    field as the generating functional for the
$ S $-matrix,
    whereas the Schwinger functional of external current, generating the
    Green functions, needs LSZ reduction formulas to define
$ S $-matrix.
    Moreover, the latter functional is not manifestly gauge invariant.

    Ironically the standard description of the background field method
\cite{BFM1},
\cite{BFM2},
\cite{Abbott}
    uses the external current and Legendre transformation.
    Alternative formulation began in
\cite{ASF}
    and entering
\cite{FS}
    was improved in
\cite{EA}.
    In what follows I use the latter approach.

    An effective action
$ W(B) $
    is a functional of a classical Yang-Mills field
$ B_{\mu}(x) $
    given by a series in a dimensionless coupling
$ \alpha $
\begin{equation*}
    W(B) = \frac{1}{\alpha} W_{-1}(B) + W_{0}(B) + \sum_{n\geq 1} \alpha^{n} W_{n}(B) ,
\end{equation*}
    where
$ W_{-1}(B) $
    is a classical action,
$ W_{0}(B) $
    is a one loop correction, defined via the determinants of vector
    and scalar operators
$ M_{1} $ and
$ M_{0} $
\begin{equation*}
    M_{0} = \nabla_{\mu}^{2} , \quad
    M_{1} = \nabla_{\sigma}^{2} \delta_{\mu\nu} + 2 [F_{\mu\nu}, \cdot] ,
\end{equation*}
    where
\begin{gather*}
    \nabla_{\mu} = \partial_{\mu} + B_{\mu} , \\
    F_{\mu\nu} = \partial_{\mu} B_{\nu} - \partial_{\nu} B_{\mu}
	+ [B_{\mu}, B_{\nu}] .
\end{gather*}
\begin{equation*}
    W_{0} = -\frac{1}{2} \ln \det M_{1} + \ln \det M_{0}
\end{equation*}
    and
$ W_{k} $,
$ k = 1,2 \ldots $
    are defined as the contribution of strongly connected vacuum diagramms
    with
$ k+1 $
    loops, constructed via Green functions
$ M_{0}^{-1} $ and
$ M_{1}^{-1} $
    and vertices defined by the forms of vector and scalar fields
$ a_{\mu}(x) $,
$ \bar{c}(x) $,
$ c(x) $
\begin{align*}
    \Gamma_{3}(B) & = g \int \tr \nabla_{\mu} a_{\nu} [a_{\mu}, a_{\nu}]
        d^{4}x ,\\
    \Gamma_{4}(B) & = g^{2} \int \tr [a_{\mu}, a_{\nu}]^{2} d^{4}x ,\\
    \Omega(B) & = g \int \tr \nabla_{\mu} \bar{c} [a_{\mu}, c] d^{4}x ,
\end{align*}
    where
$ g=\frac{1}{2}\sqrt{\alpha} $,
    taking into account anticommuting propeties of ghosts
$ \bar{c}(x) $,
$ c(x) $.

    The divergences of the diagramms should be regularized. I believe
    that there exists a regularization, defined by the cut-off momentum
$ \Lambda $
    such that all infinities are powers in
\begin{equation*}
    L = \ln \frac{\Lambda}{\mu} ,
\end{equation*}
    where
$ \mu $
    is some normalization mass. Unfortunately presently I do not know
    a satisfactory procedure for such a regularization. That is why I
    call my exposition a ``scenario''.

    The renormalizability of the Yang-Mills theory means, that there
    exists a dependence of the coupling constant
$ \alpha $ on cutoff
$ \Lambda $
    such that the full action
$ W(B) $
    is finite.

    In the case of one loop everything is clear. The functional
$ W_{0}(B) $
    can be defined via the proper time method of Fock
\cite{Fock}
    giving formula
\begin{equation*}
    W_{0}(B) = \int_{0}^{\infty} \frac{ds}{s} T(B,s) ,
\end{equation*}
    where the functional
$ T(B,s) $
    has the following behavior for small
$ s $
\begin{equation*}
    T(B,s) = T_{0}(B) + s T_{1}(B,s) 
\end{equation*}
    and
\begin{equation*}
    T_{0}(B) = \frac{1}{2} \beta_{1} W_{-1} ,
\end{equation*}
    where
$ \beta_{1} $
    is a famous negative constant.
    So the only divergence is proportional to the classical action and
    can be compensated by the renormalization of the coupling constant
$ \alpha $.
    In more detail, we regularize
$ W_{0}(B) $
    as follows
\begin{equation*}
    W_{0}^{\text{reg}}(B) = \int_{0}^{1/\Lambda^{2}} ds \, T_{1}(B,s)
	+ \int_{1/\Lambda^{2}}^{\infty} \frac{ds}{s} T(B,s)
\end{equation*}
    and choose the dependence of the coupling constant
$ \alpha $
    on
$ \Lambda $ as
\begin{equation*}
    \frac{1}{\alpha(\Lambda)} = -\beta_{1} \ln \frac{\Lambda}{m} ,
\end{equation*}
    where
$ m $
    is a new parameter with the dimension of mass.
    It is clear, that the regularized one loop
$ W(B) $
    does not depend on
$ \Lambda $
    and so is finite.
    More explicitly, we rewrite
$ W_{0}^{\text{reg}}(B) $
    as follows
\begin{equation*}
    W_{00}(B) = W_{00} + W_{01} L ,
\end{equation*}
    where
\begin{equation*}
    W_{0}^{\text{reg}}(B) = \int_{0}^{1/\mu^{2}} ds \, T_{1}(B,s)
	+ \int_{1/\mu^{2}}^{\infty} \frac{ds}{s} T(B,s) ,
\end{equation*}
    and
\begin{equation*}
    W_{01} = \beta_{1} W_{-1}
\end{equation*}
    and define the renormalized running coupling constant
\begin{equation*}
    \frac{1}{\alpha(\mu)} = - \beta_{1} \ln \frac{\mu}{m} ,
\end{equation*}
    so that
\begin{equation*}
    W_{\text{1loop}}^{\text{reg}}(B) = \frac{1}{\alpha(\mu)} W_{-1} + W_{00} .
\end{equation*}
    Thus we have traded the dimensionless
$ \alpha $
    for a dimensional parameter
$ m $.
    However
$ m $
    enters trivially, defining the scale.
    Observe that
$ \alpha(\Lambda) $ and
$ \alpha(\mu) $
    are the values of the same function for two values of the argument.
    This function satisfies the first approximation to Gell-Mann Low equation
\begin{equation*}
    x \frac{d}{dx} \alpha(x) = \beta(\alpha) = \beta_{1} \alpha^{2} ,
\end{equation*}
    where the RHS does not depend on
$ x $
    and
$ m $
    plays the role of the conserved integral.
    The shift of
$ x $
    can be interpreted as an abelian group action, this is the famous
    renormalization group.
    The same is true for the whole one loop functional
\begin{equation*}
    W_{\text{1loop}}^{\text{reg}}(B,\Lambda) = W_{\text{1loop}}(B,\mu) 
\end{equation*}
    and the renormalized action
$ W(B,\mu) $
    does not depend on the running momentum
$ \mu $.

    My scenario is based on the assumption that the whole
$ W(B) $
    has the same properties.
    I believe that all infinities in
$ W(B) $
    can be combined into the form
\begin{align*}
    W(B,\Lambda) = & \frac{1}{\alpha} W_{-1} + W_{00} + W_{01} L + \\
    & + \alpha \bigl(W_{10} + W_{11}L \bigr) + \ldots +\\
    & + \alpha^{n} \bigl(W_{n0} + W_{n1}L + \ldots
	W_{nn}L^{n} \bigr) + \ldots,
\end{align*}
    where the functionals
$ W_{k0}(B) $,
$ k=0,\ldots $
    are finite and depend on
$ \mu $.
    The coupling constant
$ \alpha(\Lambda) $
    should satisfy the full Gell-Mann Low equation
\begin{equation*}
    \Lambda \frac{d\alpha}{d\Lambda} = \beta(\alpha) =
	\beta_{1}\alpha^{2} + \beta_{2}\alpha^{3} + \ldots \beta_{n}\alpha^{n}
	    + \ldots
\end{equation*}
    and the full
$ W(B) $
    should be independent of 
$ L $
    and so it is finite.

    The equation
\begin{equation*}
    \frac{dW}{dL} = 0
\end{equation*}
    immediately gives
\begin{equation*}
    W_{11} = \beta_{2} W_{-1}
\end{equation*}
    and leads to an equation, expressed via double series in powers of
$ \alpha $ and 
$ L $.
    The condition that the corresponding coefficients vanish leads to the
    recurrent relations expressing
$ W_{nm} $
    via
$ W_{n-1,m} $,
$ n=2,3,\ldots $,
$ n\geq m $.

    Here are examples of such equations at the lowest orders
\begin{align*}
    \beta_{1}W_{10} & + W_{21} = \beta_{3} W_{-1} \\
    \beta_{2} W_{10} & + 2\beta_{1}W_{20} + W_{31} = \beta_{4} W_{-1} \\
    \beta_{1} W_{11} & + 2 W_{22} = 0 \\
    \beta_{2} W_{11} & + 2\beta_{1}W_{21} + 2W_{32} = 0 .
\end{align*}
    One can solve some of these equations exactly. For instance for the highest
    coefficients
$ W_{nn} $
    we get relation
\begin{equation*}
    (n-1)\beta_{1} W_{n-1,n-1} + nW_{nn} = 0
\end{equation*}
    and as
$ W_{11} $
    is proportional to
$ W_{-1} $
    the same is true for all
$ W_{nn} $.
    Thus their contribution
$ \sum \alpha^{n} L^{n} W_{nn} $
    is summed up to
\begin{equation*}
    \frac{\beta_{2}}{\beta_{1}} \ln (1+\beta_{1}\alpha L) W_{-1}
\end{equation*}
    and this gives the next correction to the coefficient in front of
$ W_{-1} $
    in
$ W(B) $,
    namely
\begin{equation*}
    \frac{1}{\alpha} + \beta_{1}L + \frac{\beta_{2}}{\beta_{1}}
	[\ln\alpha + \ln (\frac{1}{\alpha} + \beta_{1}L)]
\end{equation*}
    leading to the next approximation for the renormalizaed coupling
    constant
\begin{equation*}
    \frac{1}{\alpha_{r}} = \frac{1}{\alpha} + \beta_{1}L
	+ \frac{\beta_{2}}{\beta_{1}} \ln L ,
\end{equation*}
    consistent with the Gell-Mann Low equation.

    We see that all coefficients
$ W_{nm} $,
$ n\geq m $,
$ n\geq 1 $
    are expressed via the finite ones
$ W_{n0} $.
    More detailed investigation of these equations
\cite{DFI}
    shows, that the full expression for
$ W(B) $
    can be rewritten in terms of
$ W_{n0} $
    and powers of renormalized coupling constant
\begin{equation*}
    W(B,\mu) = \frac{1}{\alpha_{r}} W_{-1} + W_{00} + \sum \alpha_{r}^{n}
	W_{n0} .
\end{equation*}
    This is consistent with the equation
\begin{equation*}
    W_{\text{reg}}(B,\Lambda) = W(B,\mu)
\end{equation*}
    and
\begin{equation*}
    L|_{\Lambda=\mu} = 0 .
\end{equation*}

    Let us comment, that
    the recurrence relations make sense when the only nonzero coefficients in
$ \beta $-function
    are 
$ \beta_{1} $
    and
$ \beta_{2} $.
    This makes possible to speculate, that there exist the regularizations
    in which
$ \beta_{3} = \beta_{4} = \ldots = 0 $.

\section*{Acknowledgments}
    This work is partially supported by
RFBR grant 12-01-00207
    and the programme ``Mathematical problems of nonlinear dynamics'' of RAS.
    I thank S.~Derkachov and A.~Ivanov for collaboration in development
    of my programm.

\end{document}